\documentclass[a4paper,12pt]{article}

\usepackage{amsmath}
\usepackage{amssymb}
\usepackage{latexsym}
\usepackage{cite}
\usepackage[dvips]{graphicx}
\usepackage{epsfig}
\usepackage{amsfonts}

\newcommand{\beqa}{\begin{eqnarray}}
\newcommand{\eeqa}{\end{eqnarray}}
\newcommand{\bea}{\begin{eqnarray}}
\newcommand{\eea}{\end{eqnarray}}

\newcommand{\be}{\begin{equation}}
\newcommand{\ee}{\end{equation}}
\newcommand{\half}{\frac{1}{2}}
\newcommand{\tr}{\,\textup{tr}}

\newcommand{\mcA}{{\mathcal A}}

\newcommand{\mcF}{{\mathcal F}}

\newcommand{\mcL}{{\mathcal L}}
\newcommand{\mcN}{{\mathcal N}}

\newcommand{\mcR}{{\mathcal R}}

\newcommand{\mcW}{{\mathcal W}}
\newcommand{\mcK}{{\mathcal K}}
\renewcommand{\Re}{\textup{Re}}

\numberwithin{equation}{section}

\begin{document}
\title{\bf Moduli stabilization \\
in heterotic M-theory}
\author{\normalsize{Filipe Paccetti Correia$^{\spadesuit}$\footnote{paccetti@fc.up.pt} \ and Michael G. Schmidt$^{\clubsuit}$\footnote{m.g.schmidt@thphys.uni-heidelberg.de}},\\
{$^{\spadesuit}${\it \small{Centro de F\'\i sica do Porto}}},
\\ {{\it \small{Faculdade de Ci\^encias da Universidade do Porto}}},
\\ {{\it \small{Rua do Campo Alegre, 687, 4169-007 Porto, Portugal}}}\\
{$^{\clubsuit}${\it \small{Institut f\"ur Theoretische Physik}}},
\\ {{\it \small{Universit\"at Heidelberg}}},
\\ {{\it \small{Philosophenweg 16, 69120 Heidelberg, Germany}}}}

\date{\normalsize{August 28, 2007}}

\maketitle

\abstract{We reconsider the ingredients of moduli stabilization in heterotic M-theory. On this line we close a gap in the literature deriving the  K\"ahler potential dependence on vector bundle moduli and charged matter. Crucial in this derivation is our superspace formulation of 5d heterotic M-theory taking into account the Bianchi identities modified by brane terms. Likewise, we obtain the Fayet-Iliopolous terms due to brane localised anomalous U(1)'s. After assembling perturbative and non-perturbative contributions to the superpotential, we study supersymmetric (adS) vacua. It is found that the susy condition decouples the bundle moduli from the geometric moduli. We show that M-theory supersymmetric vacua without five-branes can be found, albeit not at phenomenologically interesting values of the geometric moduli. This result is fairly independent of the choice of vector bundle at the observable brane.}

\section{Introduction}

The last few years witnessed a notable activity, following the work of KKLT \cite{Kachru:2003aw}, in the construction of stable de Sitter vacua in type II string theory, less so in heterotic string/M-theory. One should recall, however, that the heterotic string is rather appealing in regard to particle model building, being the ideal context for the construction of supersymmetric Grand Unified Theories. Moreover, its strongly coupled limit, heterotic M-theory \cite{Horava:1995qa,Horava:1996ma}, has additional attractive phenomenological features \cite{Witten:1996mz,Banks:1996ss}, including the fact that there is an intermediate regime where the theory looks effectively five dimensional \cite{Lukas:1998yy,Lukas:1998tt}.

The proposal of KKLT \cite{Kachru:2003aw} for moduli stabilization, following the work of \cite{Giddings:2001yu}, and improvements thereof \cite{Burgess:2003ic,Saltman:2004sn,Balasubramanian:2004uy,Choi:2004sx,Bobkov:2004cy,Balasubramanian:2005zx,vonGersdorff:2005bf,Berg:2005yu,Achucarro:2006zf,Parameswaran:2006jh,Lebedev:2006qq,Dudas:2006vc,Westphal:2006tn} have been implemented mainly for the type II string. In fact, although the problem of moduli stabilization in heterotic M-theory has been object of some interest in recent years \cite{Curio:2001qi,Curio:2003ur,Buchbinder:2003pi,Becker:2004gw,Buchbinder:2004im,Anguelova:2005jr,Braun:2006th,Curio:2006dc}, it is fair to say that it is far from being a settled issue. It is well known that sometimes, in the phenomenologically interesting \emph{regime} we are forced to approach the limit where warping along the $S^1/Z_2$ direction is considerable. Then, we have to face several obstacles. Firstly, there is a constraint on the validity range of the 11d supergravity approximation. Secondly, the reduction of this theory to 4d is complicated by the fact that the different zero-modes localise differently along the bulk. Moreover, generally the zero-mode wave-functions are not explicitely known. On the other hand, the low-energy gauge theory is determined by the VEVs of the vector bundle moduli and charged scalars, all of them localised at the boundary branes. The effect of these fields in moduli stabilisation must also be taken into account.

One of the problems with these brane chiral multiplets is the poor understanding of their contribution to the K\"ahler potential. It is one of the aims of this paper to improve on this situation. The reason the K\"ahler potential dependence on vector bundle moduli $\phi$ and charged scalars $C$ is unknown is twofold. Firstly, the dependence of the gauge bundle on $\phi$ and $C$ is in general not known as this would in principle require an explicit knowledge of the solutions of the zero-slope Hermitian Yang-Mills equations on the Calabi-Yau three-fold. Unfortunatly, this seems difficult, if not impossible to find. Secondly, even if we would know the gauge bundle explicitely we lack sofar an explicit expression (i.e. in terms of holomorphic coordinates) for the K\"ahler potential dependence on the gauge bundle. It is this gap that we will be able to close in this paper. We will show in section 3 that the K\"ahler potential takes a form which is strongly reminiscent of the DeWolfe-Giddings K\"ahler potential \cite{DeWolfe:2002nn} describing a mobile D3 brane in type IIB compactifications, and that the bundle moduli and charged scalars enter the K\"ahler potential through certain real functions $k^I$, $I=1,\cdots, h^{1,1}$, dubbed "little" K\"ahler potentials, which have the form $\int_{CY} \tr(\mcA\wedge{\bar\mcA)}\wedge\omega^I$.

While in general this only provides an \emph{implicit} knowledge of the way the K\"ahler potential depends on $\phi$ and $C$, the formal expressions found in section 3 can be used to derive a number of interesting results and, in particular, study supersymmetric stable vacua of heterotic M-theory. This is due to the fact, shown in section 6, that the F-flatness conditions decouple the bundle moduli from the geometric moduli, even in the presence of brane instantons and gaugino condensates. In other words, the value of the gauge bundle in the supersymmetric vacuum is exactly \emph{independent} of the value of the geometric moduli in contrast with the expectations raised in the literature. Then, the remaining supersymmetry conditions are used to determine the values of the geometric moduli $S$ and $T^I$ in the background of the gauge bundle. By solving these equations with $h^{1,1}=1$, we will see in section 5 that the effect of membrane instantons wrapping holomorphic cycles dominates over the effect of gaugino condensation at the hidden brane in such a way that the theory is driven to an phenomenologically uninteresting limit with a far too small size of the M-theory orbifold direction, implying a too small 4d Planck mass.

In this work we will also briefly discuss the relevance for moduli stabilization in dS vacua of potentials induced by Fayet-Iliopoulos terms due to anomalous U(1)s acting on the boundary branes of heterotic M-theory. It has been recently emphasized \cite{Blumenhagen:2005ga,Blumenhagen:2006ux,Tatar:2006dc} that such U(1)'s are generically present in attempts to obtain \emph{realistic} heterotic model-building using complex line-bundles. We point out in section 5 that FI D-terms and membrane instantons in general exclude each other in the following sense. Given a K\"ahler modulus associated with an isolated holomorphic curve, either the corresponding membrane instanton superpotential vanishes or its anomalous U(1) charge vanishes.

We will perform the derivation of the bundle moduli and matter dependent K\"ahler potentials using the $N=1$ superfield description of 5d heterotic M-theory we developed in \cite{Correia:2006pj}. This is a particular application of a general superfield formalism constructed for 5d gauged supergravity on the $S^1/Z_2$ orbifold in refs.\cite{PaccettiCorreia:2004ri,Correia:2004pz,Correia:2006pj} and \cite{Abe:2004ar,Abe:2006eg}. Extending the results of \cite{Correia:2006pj} to the non-standard embedding will require a careful study of certain Bianchi identities and their dependence on the brane localised gauge bundles. It is worth emphasizing that, as we will see below, the superfield formalism turns out to be a powerfull and economic technic and is crucial for obtaining the results of this paper.

This article is organized as follows. Section 2 reviews the relevant couplings of 5d heterotic M-theory and its reduction to four dimensions. A brief explanation of the superfield description used herein is also presented. We derive explicitely the K\"ahler potentials in 3 situations. One with $h^{1,1}=1$; one with $h^{1,1}=3$; and $h^{1,1}=1$ with a mobile five-brane in the bulk. But we also discuss the structure of the K\"ahler potential in the general case. This discussion is completed in Appendix A. Section 3 contains the main technical results of this paper. We derive here the brane (and bulk-brane) superfield couplings involving the gauge bundle degrees of freedom, beyond the standard embedding. This allows us to obtain the effective K\"ahler potential dependence on the bundle moduli $\phi$ and charged matter $C$. We draw some parallels with type IIB compactifications with mobile D3-branes. In section 4 we discuss Fayet-Iliopoulos D-terms induced by anomalous U(1)'s and discuss the D-term potential in the strongly coupled case. Section 5 assembles perturbative and non-perturbative superpotentials and discusses the compatibility of the latter with non-vanishing FI terms. Finally, in section 6, we present a study of supersymmetric vacua without mobile five-branes. We conclude with a brief discussion.

\section{5d heterotic M-theory and its 4d reduction}

In the following paragraphs we review the relevant couplings of 5d heterotic M-theory. For this purpose, we use the (conformal) superfield formalism of \cite{PaccettiCorreia:2004ri,Correia:2004pz,Correia:2006pj} and \cite{Abe:2004ar,Abe:2006eg}, based on the off-shell 5d supergravity construction of \cite{Kugo:2000hn,Kugo:2000af,Fujita:2001bd,Kugo:2002js}. The point of using the superfield description is that the reduction to 4d, that we will perform below, is rather transparent, specially regarding the $\mcN=1$ conformal superspace structure.

\subsection{Relevant couplings of 5d heterotic M-theory}

As any five-dimensional gauged supergravity, 5d heterotic M-theory contains in addition to the gravitational sector a number of vector and hypermultiplets \cite{Lukas:1998yy,Lukas:1998tt}. The bosonic components of the vector multiplets are $h^{1,1}$ K\"ahler structure moduli $\{b^I\}$ and gauge fields $\{A^I\}$ obtained by expanding the 11d 3-form potential $C_3$ over (a basis of) harmonic $(1,1)$ forms. On the other hand, the hypermultiplets include the "dilaton" universal hypermultiplet and $h^{2,1}$ complex structure moduli. We will not deal here with the complex structure moduli in much detail.

As we just mentioned, the K\"ahler moduli are associated with 5d vector multiplets. Each of these decomposes into a chiral superfield $\Sigma^I$ and a vector superfield $V^I$ of $N=1$ (4d) supersymmetry as follows \cite{Kugo:2002js,PaccettiCorreia:2004ri,Abe:2004ar}:\footnote{The ellipsis in the following expressions stand for the fermionic and auxiliary components of the corresponding superfields. For a discussion of these terms see \cite{Kugo:2002js,PaccettiCorreia:2004ri,Abe:2004ar}.}
\be
                 \Sigma^I=\half e^5_y \,b^I+\frac{i}{2} A_y^I+\cdots
\ee
\be
                V^I=-\theta\sigma^{\mu}\bar{\theta} A_{\mu}^I+\cdots
\ee
Note that we take the Wess-Zumino gauge. The scalars $b^I$ have to satisfy the contraint \cite{Lukas:1998tt,Kugo:2002js}
\be\label{eq:norm}
               \mcN(b)\equiv d_{IJK}b^Ib^Jb^K=1 \ .
\ee
Note, however, that the corresponding chiral superfields $\Sigma^I$ are not constrained \cite{PaccettiCorreia:2004ri}, as they depend on one additional degree of freedom, $e^5_y$.

There is a combination of the above superfields that will play an important \emph{r\^ ole} in this paper. This is the gauge \emph{invariant}
\be\label{eq:1st_def_V_y}
                V_y^I\equiv \Sigma^I+\bar{\Sigma}^I-\partial_yV^I= e^5_y b^I+\theta\sigma^{\mu}\bar{\theta}F_{y\mu}^I+\cdots
\ee
where $F_{y\mu}=\partial_yA_\mu-\partial_\mu A_y$. Clearly, $V_y^I$ is invariant under the gauge transformations
\be
               V^I\to V^I+\Lambda^I+\bar{\Lambda}^I,\qquad\Sigma^I\to\Sigma^I+\partial_y \Lambda^I \ .
\ee
For the upcoming discussion, it is crucial to stress the fact that $F_{y\mu}$ is a component of $V_y^I$. In fact, it is well known that the brane localised gauge fields lead to modifications of the 2-form $F^I$,
\be
               F^I=dA^I-\delta(y)(\textup{brane fields}) \ .
\ee
(The precise form of the brane term will be presented in section \ref{sec:brane_localised}.) We will have therefore to modify Eq.\eqref{eq:1st_def_V_y} in such a way that $V_y^I\sim\theta\sigma^{\mu}\bar{\theta}F_{y\mu}^I$ still holds. This will be done in section \ref{sec:brane_localised}.

The crucial ingredient in 5d heterotic M-theory is the \emph{universal} hypermultiplet. It consists of four real scalars and their supersymmetric partners., and can be decomposed into two chiral $N=1$ superfields, $S$ and $\Phi$. The so-called \emph{dilaton} superfield $S$ is associated with the volume of the internal Calabi-Yau 3-fold measured at the position $y$ along the orbifold. Its lowest component is
\be\label{eq:def_S}
              S(y)=V_{CY}(y)+|\xi|^2+i\sigma \ .
\ee
Here, $\sigma$ is dual to the 5d 3-form potential ${\tilde C}_3$ arising from the 11d bulk in the Ho\v rava-Witten setup, while $\xi$ is an $S^1/Z_2$-odd complex field which appears in the expansion of the 11d 3-form potential $C_3$ (of the 4-form $G$) in harmonic forms,
\be
              C_3={\tilde C}_3+\xi\Omega+\bar{\xi}\bar{\Omega}+V_{CY}^{\frac{1}{3}}A^I\wedge\omega_I+\cdots
\ee
Finally, let us mention that $\Phi\propto\xi$ is an $S^1/Z_2$-odd chiral superfield.

According to \cite{Lukas:1998yy,Lukas:1998tt}, the 11d G-flux induces a gauging of the shift symmetry $\sigma\to\sigma+\lambda$ by a suitable combination of the 5d vector multiplets:\footnote{This shift symmetry is a particular subgroup of the isometries of the universal hyperscalar manifold. As such, from the viewpoint of 5d supergravity, this is not the most general gauging (see \cite{Fujita:2001bd} and references therein).}
\be
                     S\to S-2\epsilon(y)\alpha_I\Lambda^I \ ,
\ee
\be
                    V^I\to V^I+\Lambda^I+\bar{\Lambda}^I,\quad \Sigma^I\to\Sigma^I+\partial_y\Lambda^I \ .
\ee
The 5d superspace Lagrangian will have to be invariant under these transformations. As it was explained in \cite{Correia:2006pj}, the relevant bulk couplings, i.e. those couplings leading to non-vanishing 4d moduli interactions, can be written in the form of a D-term Lagrangian in five dimensions
\be\label{eq:master5d}
               \mcL_D=-3\int d^4\theta \,\bar{\phi}\phi \left[\mcN(V_y)\right]^{\frac{1}{3}}\, \left[S_0+\bar{S}_0+2\epsilon(y)\alpha_IV^I-2\alpha_I\int_0^y\epsilon(y)(\Sigma^I+\bar{\Sigma}^I)\right]^{\frac{1}{3}} \ ,
\ee
where $\mcN(V_y)\equiv d_{IJK}V_y^IV_y^JV_y^K$ is determined by the CY intersection numbers $d_{IJK}$ appearing also in \eqref{eq:norm}. We note that at energies below the compactification scale, the superfields $\phi$ and $S_0$ can be replaced by their zero modes, which have no $y$-dependence. In fact, as we reduce to 4d $\phi$ becomes the chiral compensator superfield and $S_0$ is the 4d chiral superfield defined by $S_0\equiv S(y=0)$.

\vspace{12pt}

The superspace 5d Lagrangian above was derived in \cite{Correia:2006pj} from a 5d supergravity perspective. Here, we will give a bottom-up way of understanding this expression. In the limit of vanishing $G_4$-flux there is no warping and the low energy 4d tree-level Lagrangian is just
\be\label{eq:unflux4d}
               \mcL_{D,4d}=-3\int d^4\theta \,\bar{\phi}\phi\exp(-\mcK/3) \ ,
\ee
with K\"ahler potential
\be
               \mcK=-\ln(S+\bar{S})-\ln\mcN(\Sigma+\bar{\Sigma}) \ .
\ee
Now, consider uplifting this expression to 5d still without $G$-flux (and warping). We must care about 5d Poincar\' e and gauge invariance. As we said above, in 5d $\Sigma^I+\bar{\Sigma}^I$ is not a gauge invariant, while $V_y^I=\Sigma^I+\bar{\Sigma}^I-\partial_yV^I$ is. Therefore, in going to 5d we have to modify \eqref{eq:unflux4d} as
\be\label{eq:unflux5d}
               \mcL_{D,4d}\to\mcL_{D,5d}=-3\int d^4\theta \,\bar{\phi}\phi \left[\mcN(V_y)\right]^{\frac{1}{3}}\, \left(S+\bar{S}\right)^{\frac{1}{3}}+\cdots \ ,
\ee
where the ellipsis stand for odd superfields which vanish in the low-energy limit. Of course, 5d Poincar\' e invariance must be also imposed both in the vector and dilaton sector. We will here not care about the vector sector, see \cite{Correia:2006pj} for details. The only way of recovering 5d Poincar\'e invariance in the dilaton sector is to introduce an F-term Lagrangian which must have the form (see e.g. \cite{Bagger:2006hm})
\be
                 \mcL_{F,5d}=\int d^2\theta\,\phi^3\,\Phi\,\partial_y S +\textup{h.c.}+\cdots,
\ee
where $\Phi$ is an odd chiral superfield\footnote{In the notation of \cite{Correia:2006pj}, $\Phi$ is an odd combination of several superfields defined therein: $4\Phi=h_2^c/h_1+H^c(h_1+H)/(h_1h_2)$.}.

Let us now turn on the $G$-flux. We know that this correponds to gauging the shift symmetry $S\to S+i\lambda$ by a suitable combination of the 5d vector multiplets. Imposing invariance of $\mcL_{F,5d}$ under the action of this transformation we find that $\partial_y S$ must be replaced by $\partial_y S+2\epsilon(y)\alpha_I\Sigma^I$. Likewise, in $\mcL_{D,5d}$ we must replace ($S+\bar{S}$) by ($S+\bar{S}+2\epsilon(y)\alpha_IV^I$). Finally, we introduce $S_0$ as
\be\label{eq:S_0}
                     S=S_0-2\alpha_I\int_0^y\epsilon(y)\Sigma^I \ .
\ee
To fully understand the meaning of $S_0$ we must note that now $\mcL_{F,5d}$ reads
\be
                      \mcL_{F,5d}=\int d^2\theta\,\phi^3\,\Phi\,\partial_y S_0 +\textup{h.c.}+\cdots
\ee
Going to 4d the odd superfield $\Phi$ acts as multiplier imposing that  $\partial_y S_0=0$ \cite{Correia:2006pj,Abe:2006eg}. This means that $S_0$ is a 4d superfield and according to \eqref{eq:S_0} it is the value of the dilaton measured at the visible brane ($y=0$). Finally, plugging \eqref{eq:S_0} back into \eqref{eq:unflux5d} we obtain the D-term Lagrangian \eqref{eq:master5d}.

\subsection{Reduction to 4d and the effective K\"ahler potential}\label{sec:reduction_without_branes}

In the preceding discussion we have seen how starting from the known 4d weakly coupled heterotic theory, one can uplift to 5d and turn on the G-flux to obtain the superspace Lagrangian \eqref{eq:master5d}. We will now go the opposite way from 5d to 4d, but keeping the non-vanishing warping. It is not difficult to see that the 4d K\"ahler potential is obtained by performing the following integral (cf. \eqref{eq:unflux4d})
\be\label{eq:master_for_kaehler}
                e^{-\mcK/3}=\oint dy\,\left[\mcN(\Sigma+\bar{\Sigma})\right]^{\frac{1}{3}}\, \left[S_0+\bar{S}_0-2\alpha_I\int_0^y\epsilon(y)(\Sigma^I+\bar{\Sigma}^I)\right]^{\frac{1}{3}},
\ee
where we retained only $S^1/Z_2$-even quantities. The difficulty now is to evaluate this expression explicitely in the warped background. As it was pointed out in \cite{Correia:2006pj}, it is only in the $h^{1,1}=1$ case that this is straightforward. We obtain \cite{Lalak:2001dv,Correia:2006pj}
\be\label{eq:K_potential_universal}
               \mcK=-3\ln\frac{3}{4\alpha}\left[(S_0+\bar{S}_0)^{\frac{4}{3}}-(S_\pi+\bar{S}_\pi)^{\frac{4}{3}}\right],
\ee
where the modulus $S_{\pi}\equiv S_0-\alpha T$ measures the size of the CY 3-fold at the location of the hidden brane, and\footnote{ In this paper, the definition of the moduli $T^I$ differs from the one in \cite{Correia:2006pj} by a factor of $2\pi$.}
\be
                T\equiv\oint \, dy\Sigma = \int_0^{\pi} dy e^5_y + i\int_0^{\pi} dy A_y +\cdots
\ee
In the weak coupling limit $\alpha\Re(T)\ll\Re(S)$ we recover the universal K\"ahler potential
\be\label{eq:weak_kaehler}
                \mcK=-\ln(S+\bar{S})-3\ln(T+\bar{T})
\ee
where we introduced the \emph{average} CY volume modulus $S\equiv (S_0+S_\pi)/2$. Note that due to warping, in the strong coupling regime the nice direct product of the hypermultiplet sector and the vector sector patent in \eqref{eq:weak_kaehler} disappears. Moreover, unlike the weak coupling K\"ahler potential \eqref{eq:weak_kaehler}, the exact K\"ahler potential \eqref{eq:K_potential_universal} describes a moduli space which has a curvature singularity at the strong coupling limit $\textup{Re}(S_\pi)=0$.

\vspace{12pt}

Before considering the addition of bulk $M5$ branes, let us comment on the structure of the K\"ahler potential in the more general $h^{1,1}>1$ case. The first fact we want to stress is that $\mcK$ is now a function of $\Re(S_0)$ and of the 4d K\"ahler moduli $\Re(T^I)$, where
\be
                T^I\equiv\oint \, dy\Sigma^I ,
\ee
and, again, $\Re(S_0)$ measures the size of the CY 3-fold at the visible brane. Note that, instead of, say, $T^0$ one can use the modulus $S_\pi\equiv S_0-\alpha_I T^I$ which sets the size of the CY at the hidden brane, or the average CY size modulus $S$ defined above. As a non-trivial example, let us consider the STU-model with $h^{1,1}=3$, $\alpha_0\neq 0$, $\alpha_1=\alpha_2=0$ and $d_{123}=1$. The K\"ahler potential can be determined using the methods of \cite{Correia:2006pj,Abe:2006eg} and turns out to be
\be
               \mcK_{STU}=-\ln\frac{1}{2\alpha_0}\left[(S_0+\bar{S}_0)^2-(S_\pi+\bar{S}_\pi)^2\right]-\ln(T^1+\bar{T}^1)(T^2+\bar{T}^2) .
\ee

It is important to note that in the two examples presented in this section, where we are able to explicitely calculate the low-energy K\"ahler potential, the following set of relations is satisfied:
\be
              \mcK_i(X^i+\bar{X}^i)=-4,
\ee
\be
              \mcK_{i\bar{j}}(X^i+\bar{X}^i)=-\mcK_{\bar{j}},
\ee
\be
              \mcK^{\bar{j}i}\mcK_i\mcK_{\bar{j}}=4,
\ee
where the $X^i$ denote collectively the $S$ and $T$ moduli. Moreover, these expressions still hold in the presence of M5 branes in the bulk. This leads us to conjecture that they should be valid also in the general case, regardless of the explicit form of the intersection numbers $d_{ijk}$ and the number of five-branes in the 5d bulk. We show in appendix \ref{app:A} that this is true, at least for general $d_{ijk}$ and no five-branes.

\subsection{Bulk M5 branes}\label{subsec:5branes}
Sometimes it is useful to introduce bulk five-branes to more easily satisfy the tadpole cancelation conditions \cite{Lukas:1998hk,Donagi:1998xe,Donagi:1999gc,Donagi:2000zf}. For simplicity we will specialize to the case with $h^{1,1}=1$. There is therefore only one 2-cycle the five-brane can wrapp. The presence of the five-brane in the bulk modifies the $y$-dependence of the $G_4$ flux. This can be very easily accounted for by replacing the coupling $\alpha_0\epsilon(y)$ in Eq.\eqref{eq:master_for_kaehler} by $\alpha(y)\epsilon(y)$ with
\be
                \alpha(y)=\left\{\begin{array}{ll}\alpha_0 & ,0< |y|< z \\ \alpha_1 &, z< |y|< \pi\end{array}\right.
\ee
where $z$ is a \emph{fixed} reference distance of the five-brane to the visible boundary, $\int^z dy\, e_y^5$ being the physical distance.

The K\"ahler potential was determined in \cite{Correia:2006pj}. It is given by
\be
            \mcK=-3\ln\frac{3}{4}\left[\alpha_0^{-1}(S_0+\bar{S}_0)^{\frac{4}{3}}- (\alpha_0^{-1}-\alpha_1^{-1})(S_5+\bar{S}_5)^{\frac{4}{3}}- \alpha_1^{-1}(S_\pi+\bar{S}_\pi)^{\frac{4}{3}}\right] \ .
\ee
Here, $\textup{Re}(S_5)$ is the volume of the internal CY measured at the position of the five-brane, which can be rewritten in terms of the five-brane position modulus $Z$ as\footnote{The definition of the modulus $Z$ used in \cite{Correia:2006pj} differes from the present one by a factor of $2$.}
\be
                 S_5=S_0-\alpha_0 Z \ ,
\ee
where \cite{Moore:2000fs,Derendinger:2000gy,Brandle:2001ts}
\be
                 Z \equiv \int^z_0 dy\, \left( e_y^5+i A_y \right) -i B \ ,
\ee
and $B$ originates from the KK reduction of the 2-form living on the five-brane. Finally, the modulus $S_\pi$ can also be expressed in terms of the five-brane modulus $Z$ as
\be
                 S_\pi=S_0-\alpha_1 T -(\alpha_0-\alpha_1)Z \ .
\ee
We note that $\beta\equiv(\alpha_1-\alpha_0)$ is the tension of the five-brane and therefore must be positive, that is
\be
                 \alpha_1>\alpha_0 \ .
\ee

\section{Brane localised couplings and the K\"ahler potential}\label{sec:brane_localised}

The requirement of global anomaly cancelation leads necessarily to non-trivial gauge bundles and therefore to the breaking of the $E_8$ gauge symmetries at least at one of the boundary planes. The resulting low-energy brane field content as well as its interactions depend on the precise choice of the vector bundle and also on the CY 3-fold. We will see now how by solving the Bianchi identities one can derive the contribution of the brane multiplets to both the K\"ahler and superpotential. To illustrate this idea, we will first describe the brane localised couplings in the \emph{simplest} case of the standard embedding with universal moduli only. The general case of a non-standard embedding with $h^{1,1}\neq 1$ will then be discussed below.

\subsection{Standard embedding with universal moduli}\label{subsec:kaehler_brane}

In the standard embedding the gauge group at the visible brane reduces to an $E_6\subset E_8$ and, in addition to the corresponding gauge fields, there are $h^{1,1}$ charged multiplets $C^I$ transforming in the $\bf{\bar{27}}$ representation of the $E_6$ (for simplicity we assume $h^{2,1}=0$). We will consider the case $h^{1,1}=1$. Since we are dealing here with 5d supergravity, the kinetic terms for $C$ must be encoded in a D-term Lagrangian localised at the visible brane,
\be\label{eq:local_lagrange}
                 \mcL_{\textup{matter}}^D=-3\delta(y)\int d^4 \theta \bar{\phi}\phi\, \Omega(C,\,\bar{C};\,S+\bar{S};\,V_y).
\ee
This term can be determined by solving the Bianchi identity for the 5d 2-form $F$ in the presence of (a gradient of) the matter field $C$. In fact, the only natural and gauge invariant way of obtaining such a term is by performing the following addition to \eqref{eq:master5d}:
\be
               V_y\to V_y=\Sigma+\bar{\Sigma}-\partial_yV-\delta(y)\bar{C}C .
\ee
This modification is required to ensure the gauge invariance of the action and can be obtained by noting that $V_y^I=\cdots+i{\bar\theta}\sigma^\mu\theta F_{\mu y}^I$. In fact, in the presence of brane matter, the 5d Bianchi identities have brane localised sources, leading to
\be
               F=dA+\delta(y)(i\bar{C}D_\mu C+\textup{c.c.}) dx^\mu \wedge dy .
\ee

The above redefiniton of $V_y$ must be accompanied by a modification in the definition of the superfield $\Sigma$,
\be\label{eq:redef_sigma}
               \Sigma\to\Sigma=\half (e^5_y \,b+\delta(y)|C|^2)+\frac{i}{2} A_y+\cdots
\ee
The rational behind this is to impose that the lowest component of the superfield $\Omega$ in Eq.\eqref{eq:local_lagrange}, which is the coefficient of the Einstein-Hilbert term \cite{Kugo:1982mr}, has no support on the brane. In this way we ensure that no brane localised Einstein-Hilbert terms exist. Likewise, the dilaton superfield $S_0$ (see \eqref{eq:def_S}) is modified as 
\be
               S_0=V_{CY}(0)+\frac{\alpha}{2}|C|^2+i\sigma(0) \ ,
\ee
(with $\xi(0)=0$) otherwise the bulk Lagrangian would depend on the brane field $C$.

To determine the low-energy K\"ahler potential we have now to perform the following integral:
\be\label{eq:kaehler_C}
                e^{-\mcK/3}=\int dy\,\left[\mcN(\Sigma+\bar{\Sigma}-\delta(y)C{\bar C})\right]^{\frac{1}{3}}\, \left[S_0+\bar{S}_0-2\alpha\int_0^y\epsilon(y)(\Sigma+\bar{\Sigma})\right]^{\frac{1}{3}} \ .
\ee
It is enough to calculate the lowest component of this superfield expression, since the rest can be obtained by supersymmetrization \cite{Correia:2006pj}. (This is the reason we dropped the vector superfield $V$ in Eq.\eqref{eq:kaehler_C}.) The lowest component of \eqref{eq:kaehler_C} is (with $d_{111}=1$)
\be\begin{split}
               e^{-\mcK/3} & = 2\int_0^\pi dy\,e^5_y\, \left[2V_{CY}(0)-2\alpha\int_0^y e_y^5 \right]^{\frac{1}{3}}\\
                            & =\frac{3}{4\alpha}(2V_{CY})^{\frac{4}{3}}(0)-\frac{3}{4\alpha}\left(2V_{CY}(0)-2\alpha\int_0^\pi dy\,e^5_y\right)^{\frac{4}{3}} \ .
\end{split}\ee
Notice that, due to the definition of $\Sigma$, eq.\eqref{eq:redef_sigma}, we have now
\be
               T=\oint dy\Sigma=\int_0^\pi dy(e^5_y+iA_y)+\half|C|^2 \ .
\ee
We thus see that in terms of the holomorphic coordinates the K\"ahler potential reads
\be\begin{split}\label{eq:kaehler_universal_standard}
               \mcK & = \mcK(S_0+\bar{S}_0-\alpha|C|^2;\,T+{\bar T}-|C|^2)\\
                    & = -3\ln \frac{3}{4\alpha}\left[(S_0+\bar{S}_0-{\alpha}|C|^2)^{\frac{4}{3}}-(S_0+\bar{S}_0-{\alpha}(T+\bar{T}))^{\frac{4}{3}}\right]    \ .
\end{split}\ee
In the $\alpha\to 0$ limit we find the well known result
\be
               \mcK=-\ln(S+\bar{S}-{\alpha}|C|^2/2)-3\ln(T+\bar{T}-|C|^2) \ .
\ee

It is important to stress that the combination $T+\bar{T}-|C|^2$, appearing in the K\"ahler potential, is still proportional to the \emph{undeformed} volume of the 2-cycle:
\be\label{eq:undeformed_cycle}
               T+\bar{T}-|C|^2=\oint dy e^5_y \sim \oint dy\sqrt{g_{11}}\int_{2 cycle} J \ ,
\ee
where $J$ is the K\"ahler 2-form of the CY 3-fold. By undeformed volume we mean the following. To leading order, the effect of turning on the gauge bundles at the boundaries is just to produce an 11d metric which is a warped product of a CY 3-fold and a 5d spacetime. At this level of approximation, the only effect of the G-flux is the CY volume's dependence on $y$. Clearly, Eq.\eqref{eq:undeformed_cycle} gives the volume of the cycle at this level of approximation. Notice, in particular, that it does not depend on the value of $C$. In fact, this should be expected, as the warping is determined by topological numbers and it therefore cannot change continuously as we change the gauge bundles at the boundaries. On the other hand, in general we do expect the volume of the cycles to be \emph{deformed} continuously in the presence of these gauge bundles, that is, it should depend on the vector bundle moduli $\phi^\alpha$ and charged matter $C^a$. As we will argue in section \ref{sec:susy_vacua}, while the K\"ahler potential depends on the \emph{undeformed} volume of the cycles, the superpotential induced by brane instantons will be a function of the \emph{deformed} volume of the wrapped holomorphic cycles and will thus depend on $\phi^\alpha$ and $C^a$. Note that the same reasoning applies to the dilaton superfields $S_0$ and $S_\pi$ and the fact that the K\"ahler potential depends only on the \emph{warped} volume of the CY, which is sensitive only to the topology of the gauge bundle.

\subsection{Non-standard embedding}

In the general case, that is for other choices of the gauge bundle, additional states will be present at the 4d branes. In particular, in addition to the chiral multiplets $C^a$ charged under the 4d gauge groups, one finds vector bundle moduli $\phi^\alpha$ whose number depends on the topological properties of the vector bundle. The VEVs of the bundle moduli determine the configuration of the gauge bundle and thus also the value of Yukawa couplings and matter kinetic terms. As such, it is important to know how and at which values they are stabilized. As we have seen above, their kinetic terms, determined by the way they enter the 4d K\"ahler potential, can be obtained by looking at the 5d Bianchi identities. A direct reduction from the 11d Bianchi identity for the 4-form $G$ shows that
\be\label{5dfieldstrength}
               F^I=dA^I-\lambda\,\delta(y)dy\wedge dx^\mu\tr_{E_8}\left[\mcA_\mu\int_{c^I}\tilde{d}\tilde{\mcA}+\int_{CY}(D_\mu\tilde{\mcA})\wedge\tilde{\mcA}\wedge\omega^I\right],
\ee
where we decomposed the $E_8$ gauge field into a component along the CY 3-fold $\tilde{\mcA}$ and a component along the macroscopic four dimensions $\mcA=\mcA_\mu dx^\mu$. Moreover, $\tilde{d}$ is the exterior derivative operator defined on the CY while $D_\mu\equiv \partial_\mu+[\mcA_\mu,\cdot]$. The $\{c^I\}$ form a basis of $h^{1,1}$ cycles such that
\be
                  \int_{c^I}\omega_J=\int_{CY}\omega^I\wedge\omega_J=\delta^I_{J}.
\ee
Note that above and in the following, for simplicity, we suppress similar terms that can be induced by a non-trivial gauge bundle at the hidden brane.

In Eq.\eqref{5dfieldstrength} we have two source terms. The first one will be discussed in section \ref{sec:FI} as it can lead to Fayet-Iliopoulos D-terms. The second,
\be
                  \delta(y)\int_{CY}\tr_{E_8}\left[(D_\mu\tilde{\mcA})\wedge\tilde{\mcA}\wedge \omega^I\right],
\ee
is manifestly gauge invariant in respect to the \emph{unbroken} 4d gauge transformations, and it depends on the (covariant) derivative of $\tilde\mcA$. 
We introduce now a set of (local) complex coordinates along the CY 3-fold, $\{z^n\}$ ($n=1,2,3$), and write
\be
                   \tilde{\mcA}={\mcA}_ndz^n+\bar{\mcA}_{\bar n}d\bar{z}^{\bar n}.
\ee
Recalling that $\tilde\mcA=\tilde{\mcA}[\phi,\bar{\phi};C,\bar{C}]$ and the fact that $\partial_{\alpha,a}{\mcA}_{n}=0$, it follows that $F^I_{\mu y}$ is a component of the superfield $V_y^I$ where now
\be\label{superfield_source}
                V_y^I=\Sigma^I+\bar{\Sigma}^I-\partial_yV^I - \delta(y)k^I(\phi,\bar{\phi}\,;C,\bar{C} ),
\ee
and the superfield $k^I$
is found to be\footnote{In fact, we are determining $k^I$ up to a sum of a holomorphic and an anti-holomorphic term which can be shifted into the definition of $T^I$ without affecting our conclusions below.}
\be\label{eq:kappa^I}
                   k^I=i\lambda\int_{CY}\tr_ {E_8}\left(\mcA_n dz^n \wedge \bar{\mcA}_{\bar n}d{\bar z}^{\bar n}\right)\wedge\omega^I,
\ee
which can be easily checked to be real. As in the universal case discussed above, the 5d K\"ahler moduli $\Sigma^I$ are modified as
\be
                   \Sigma^I=\half (e^5_y \,b^I+\delta(y)k^I)+\frac{i}{2} A_y^I \ ,
\ee
and therefore
\be
                   T^I=\int_0^\pi dy\,(e^5_y \,b^I+i A_y^I)+\half k^I \ .
\ee
Likewise, the lowest component of the dilaton superfield also depends on the brane fields as
\be
                   S_0=V_{CY}(0)+\alpha_I k^I(\phi,{\bar\phi};C,{\bar C})+i\,\sigma(0) \ .
\ee

We can use now the same reasoning as before to show that in the general case the effective 4d K\"ahler potential reads
\be\label{eq:kaehler_general_brane_stuff}
                \mcK=\mcK\left(S_0+{\bar S}_0-\alpha_Ik^I(\phi,\bar\phi ; C,\bar{C})\,;T^I+{\bar T}^I- k^I(\phi,\bar\phi ; C,\bar{C}) \right) \ .
\ee
In the case of the standard embedding with universal moduli only we have $k=|C|^2$ and Eq.\eqref{eq:kaehler_general_brane_stuff} reduces to Eq.\eqref{eq:kaehler_universal_standard}.

As a non-trivial test of this result let us consider the weak coupling limit $\alpha_I\to 0$. In this case the K\"ahler potential is
\be
           \mcK=-\ln(S+\bar{S})+\mcK^{1,1}(T^I+\bar{T}^I- k^I)\simeq-\ln(S+\bar{S})+\mcK^{1,1}(T+\bar{T})-\mcK^{1,1}_{I}k^I \ ,
\ee
where $\mcK^{1,1}(T+\bar{T})$ is the K\"ahler potential for the K\"ahler moduli. The last term on the r.h.s. can be regarded as the brane fields K\"ahler potential and reads
\be
           \mcK_{brane}=-\mcK_{I}^{1,1}k^I=-i\lambda\int_{CY}\tr_ {E_8}\left(\mcA_n dz^n \wedge \bar{\mcA}_{\bar n}d{\bar z}^{\bar n}\right)\wedge \mcK_{I}^{1,1}\omega^I \ .
\ee
Finally, this can be rewriten as
\be\label{eq:brane_kaehler}
             \mcK_{brane}\simeq 3i\lambda\frac{\int_{CY}\tr\left(\mcA \wedge \bar{\mcA}\right)\wedge J\wedge J}{\int_{CY} J\wedge J\wedge J}\sim \frac{\lambda}{2}\tr\int\sqrt{g_{CY}}\,g^{n\bar{m}}_{CY}\mcA_n  \bar{\mcA}_{\bar m} \ ,
\ee
where $J$ is the K\"ahler 2-form of the CY 3-fold. This expression matches the K\"ahler potential that can be obtained by a direct reduction of the 10d Yang-Mills action, see e.g.\cite{Buchbinder:2003pi}. Note that also the cross term between the $T$-moduli and the bundle moduli mentioned in this reference follows directly from the above K\"ahler potential. There is no need for an additional contribution to $\mcK$ to explain this term, in contrast to the expectations expressed in \cite{Buchbinder:2003pi}.

Notice that we can also determine a one-loop correction to the bundle moduli and matter K\"ahler potential \eqref{eq:brane_kaehler}, which reads
\be\label{eq:brane_1loop}
               \mcK_{brane}^{1-loop}\simeq -\mcK_{S_0}\alpha_I k^I \sim -i\frac{\lambda}{S_0+{\bar S}_0}\int_{CY}\tr\left(\mcA \wedge \bar{\mcA}\right)\wedge\left[\tr(\mcF^2)-\half\tr(\mcR^2)\right] \ ,
\ee
and depends on the second Chern characters of the vector and tangent bundles.

\vspace{12pt}

A few remarks are now in order. Firstly, we would like to stress that the way $\mcK$ depends on the brane fields is \emph{not} unexpected. A similar expression, the DeWolfe-Giddings K\"ahler potential \cite{DeWolfe:2002nn}, has been proposed in the context of type IIB compactifications with mobile D3 branes, where it can naturally be matched to the kinetic terms derived from the Born-Infeld action for D3-branes.\footnote{A relevant difference between our case and the DeWolfe-Giddings K\"ahler potential is that the latter is only known to apply when $h^{1,1}=1$. In that case the \emph{little} K\"ahler potential $k$ is just the K\"ahler potential of the compactification CY evaluated at the position of the mobile D3 brane \cite{DeWolfe:2002nn}. It is natural to expect that this expression extends to $h^{1,1}>1$, although it is not clear what the $k^I$ will be in that case.} An important feature of our K\"ahler potential is that the vector bundle moduli cannot be disentangled from the geometric moduli. It is therefore excluded that the K\"ahler potential for the former be
\be
               -p\ln(\phi+\bar\phi) \ ,
\ee
as advocated e.g. in \cite{Buchbinder:2003pi}. Moreover, the above form for the K\"ahler potential preserves an important property already mentioned in section \ref{sec:reduction_without_branes}, namely that
\be\label{eq:again_thesame_property}
               \mcK^{i{\bar j}}\mcK_i\mcK_{\bar j}=4 \ ,
\ee
where $i,\bar{j}$ run over all moduli and matter, even in the presence of the brane fields. We prove this assertion in the appendix. This is the heterotic M-theory counterpart of the no-scale structure property in the type IIB case. 

Finally, let us comment on the structure of $k^I(\phi,\bar{\phi};~C,\bar{C})$. To do this, we note that the bundle moduli and charged matter fields are introduced as
\be
                     \bar{\mcA}=\bar{\mcA}(\phi)+C \ ,
\ee
where $\bar{\mcA}(\phi)$ is a solution of the zero-slope limit of the hermitian Yang-Mills equations, and the charged fluctuations $C$ span the Dolbeault cohomology $H^1(CY_3,V_S)$, where $V_S$ is the vector bundle in the representation $S$ at the $y=0$ brane. Let us introduce a set of 1-forms $\{{\bar u}_a\}$ forming a basis of $H^1(CY_3,V_S)$. Then, $C={\bar u}_a C^a$ and we see that $k^I$ consists of two pieces
\be
                 k^I={ k}^I_\phi(\phi,\bar{\phi})+G^I_{a\bar{b}}(\phi,\bar{\phi})\bar{C}^{\bar{b}}C^{a} \ ,
\ee
where $k^I_\phi$ depends only on the bundle moduli and is obtained by evaluating Eq.\eqref{eq:kappa^I} for solutions of the zero-slope limit of the hermitian Yang-Mills equations, while
\be\label{eq:kaehler_metric}
                 G^I_{a\bar{b}}(\phi,\bar{\phi})=i\lambda\int_{CY}\tr_{E_8}(u_{\bar b}\wedge {\bar u}_{a})\wedge \omega^I \ .
\ee
It would be interesting to evaluate these expressions explicitly, e.g. using the gauge instanton solution studied in \cite{Gray:2003vw}. We leave this for future work.

\section{Anomalous U(1)s and Fayet-Iliopoulos terms}\label{sec:FI}

Recent searches for realistic heterotic model-building using non-trivial line-bundles naturally lead to the presence of anomalous U(1)s in the gauge group content of both the visible and the hidden branes (see \cite{Blumenhagen:2005ga,Tatar:2006dc,Blumenhagen:2006ux} and references therein). The effects of such anomalous U(1)$_X$ factors was studied long ago in \cite{Lukas:1999nh} in the absence of bulk M5 branes. Their results were recently extended to include M5 branes in \cite{Blumenhagen:2006ux}. One of the facts pointed out in \cite{Blumenhagen:2006ux} is that, in the effective 4d theory, not only the dilaton but also the K\"ahler moduli and the five-brane moduli will be charged under the anomalous U(1)$_X$. We will show below how their charges relate to each other.

Important to us is the fact that the U(1)$_X$ induce moduli-dependent Fayet-Iliopolous terms, leading to positive contributions to the potential energy. As it was first pointed out in the context of type IIB flux compactifications, these FI terms have the potential of helping with uplifting possible AdS$_4$ vacua \cite{Burgess:2003ic,Achucarro:2006zf}. This possibility was also explored in the heterotic M-theory setup in ref.\cite{Buchbinder:2004im}. However, as we will argue in this section, this reference did not include all terms relevant for the analysis.

To study the effects of the anomalous U(1)$_X$ we have once again to take into account the modification to the 5d Lagrangian presented in the previous section. This modification is supported at the orbifold branes where the U(1)$_X$ is acting. We will assume, for notational simplicity, that there is just one U(1)$_X$ at the visible brane, none in the hidden one. The generalization to multiple anomalous U(1)s in both branes is straightforward. The modified Bianchi identities for the bulk gauge multiplets lead to \cite{Lukas:1999nh}
\be
                    F^I=dA^I- q^I\delta(y)\mcA_X\wedge dy \ ,
\ee
where, according to Eq.\eqref{5dfieldstrength}, the charges $q^I$ are the topological invariants
\be
                    q^I\sim-\lambda\int_{c^I}\tilde{d}\tilde{\mcA}_X=-\lambda\int_{c^I}c_1(\tilde{\mcA}_X) \ ,
\ee
determined by the first Chern class $c_1(\tilde{\mcA}_X)$ of the vector bundle. In our superspace formalism this corresponds to rewriting $V_y^I$ as
\be
                    V_y^I= \Sigma^I+\bar{\Sigma}^I-\partial_y V^I - q^I\delta(y)V_X \ ,
\ee
where $V_X$ is the 4d vector superfield (in Wess-Zumino gauge) transforming under the U(1)$_X$ as
\be
                    V_X\to V_X +\Lambda_X+\bar{\Lambda}_X \ .
\ee
Moreover, the condition that $F^I$ is a gauge invariant implies that ${V}_y^I$ has to be invariant under this transformation. Thus, we find that $\Sigma^I$ shifts under the anomalous U(1)$_X$,
\be
                    \Sigma^I\to\Sigma^I+q^I\delta(y)\Lambda_X \ .
\ee
In component notation this means that $A^I_y\to A^I_y+ q^I\delta(y)\lambda_X(x) $. This shift is necessary to cancel out the brane localised anomaly \cite{Lukas:1999nh}.

As we just pointed out, the only effect of the anomalous U(1)$_X$ on the bulk superspace Lagrangian is to rewrite $V_y$, introducing brane-localised terms. The latter induce tadpole terms for $V^I$ proportional to $V_X$, which source "non-zero" modes
\be\label{eq:nonzeromodeVX}
                   V^I=-\half q^I V_X\epsilon(y)(1-|y|/\pi)+{\tilde V}^I \ ,
\ee
leading to
\be
                   {V}_y^I=\Sigma^I+\bar{\Sigma}^I-\partial_y\tilde{V}^I-\frac{q^I}{2\pi}V_X \ .
\ee
Here, $\tilde{V}^I$ stands for the y-dependent fluctuations which, since the vector superfield $V^I$ is $S^1/Z_2$-odd, vanish at low energies. Going down to 4d, the above non-zero modes will contribute $V_X$-dependent terms to the effective K\"ahler potential. We will see now what those terms are.

To determine the effective K\"ahler potential, using Eqs.\eqref{eq:master5d} and \eqref{eq:nonzeromodeVX} we find that we have to perform the following integral
\be
                 e^{-\mcK/3}=\int dy  \left[\mcN(\Sigma+\bar{\Sigma}-\tfrac{q}{2\pi}V_X)\right]^{\frac{1}{3}}\,\left[S_0+\bar{S}_0-\alpha_I q^I V_X-2\alpha_I\int_0^y\epsilon(y)(\Sigma^I+\bar{\Sigma}^I-\tfrac{q^I}{2\pi}V_X)\right]^{\frac{1}{3}} \ ,
\ee
where we supressed all odd superfields and possible brane chiral matter contributions. Now, we know that for $V_X=0$ and no brane chiral matter, the K\"ahler potential depends only on the 4d moduli $S_0$ and $T^I$. Therefore, if $\mcK(S_0+\bar{S}_0;\,T^I+\bar{T}^I)$ is the K\"ahler potential for $V_X=0$, then we find that turning on $V_X$ the K\"ahler potential is
\be
                  \mcK=\mcK\left(S_0+\bar{S}_0-\alpha_I q^I V_X;\,T^I+\bar{T}^I-q^I\,V_X\right) \ .
\ee

To obtain the D-term potential due to the FI terms we will assume that $V_X$ is normalized such that the anomalous gauge kinetic term is
\be
                    \frac{1}{4}\int d^2\theta S_0\,W_XW_X + \textup{c.c.}
\ee
We find then
\be
                    V_D=\half\frac{(q^I\alpha_I \mcK_{S_0}+q^I\mcK_I)^2}{S_0+\bar{S}_0}.
\ee
As pointed out in \cite{Blumenhagen:2005ga}, the D-flatness condition $q^I(\alpha_I \mcK_{S_0}+\mcK_I)=0$ can be re-interpreted as a 1-loop corrected Donaldson-Uhlenbeck-Yau integrability condition. This is intimately related to the appearence of a 1-loop correction to the brane kinetic term \eqref{eq:brane_1loop}. 
To illustrate the relevance of FI D-term potentials for moduli stabilization, we study the tractable $h^{1,1}=1$ case. Using Eq.\eqref{eq:K_potential_universal} we find that the D-term potential reads
\be\label{sec:Dpotentialuniv}
                    V_D=\frac{8\alpha^2q^2}{(S_0+\bar{S}_0)^{\frac{1}{3}}}\frac{1}{\left[(S_0+\bar{S}_0)^{\frac{4}{3}}-(S_\pi+\bar{S}_\pi)^{\frac{4}{3}}\right]^2} \ .
\ee
It is interesting to consider the weak-coupling limit $\alpha\textup{Re}(T)\ll \textup{Re}(S_0)$. We find to leading order
\be
                    V_D\simeq \frac{9q^2}{2}\frac{1}{(S+\bar{S})(T+\bar{T})^2} \ ,
\ee
which clearly differs from the result found in \cite{Buchbinder:2004im}. Note that this difference is not qualitative, because the physical force towards increasing both $\textup{Re}(S)$ and  $\textup{Re}(T)$ is also present in the D-term potential of \cite{Buchbinder:2004im}. This force drives the theory to the phenomenologically interesting limit of strong coupling at the hidden brane and might be a means of obtaning (supersymmetry breaking) stable dS vacua. We will explore this possibility in a forthcoming publication.



We close this section recalling how the above reasoning is changed once we take into account the effects of matter charged under the anomalous U(1)$_X$ acting at the visible brane. We denote by $C$ the chiral multiplets with U(1)$_X$ charges $Q_C$. Note that we are suppressing non-abelian indices. The main effect of the presence of the charged matter at the visible brane is to induce the following change in the D-term
\be
               D_X\sim \left(q^I+\sum_C\, Q_C\, G^I_{C{\bar C}}(\phi,\bar{\phi})|C|^2\right)(\alpha_I \mcK_{S_0}+\mcK_I) \ .
\ee
As expected, the inclusion of the charged matter fields induces a potential that eventually can be canceled by giving suitable Vevs to the charged scalars. If this happens, then \eqref{sec:Dpotentialuniv} is of no use for moduli stabilization.

\section{Superpotential}\label{subsec:superpotential}

In this section we discuss the perturbative and non-perturbative superpotentials needed for moduli stabilization.

The perturbative superpotential can be derived in several ways. One is to use the well-known fact that in the presence of non-zero four-form flux $G$, an effective 4d superpotential \cite{Gukov:1999ya,Behrndt:2000zh,Becker:2002jj,Anguelova:2006qf}
\be\label{eq:flux_superpotential}
                      W=\int_{M_7}G\wedge{\Omega}
\ee
($M_7=CY_3\times S^1/Z_2$) is generated. We will study first the superpotential dependence on the brane fields. Solving the Bianchi identities one finds the following brane induced $G$
\be
                    G_{brane}\sim-\lambda\, dy\wedge\tr_{E_8}\left[{\tilde\mcA}\wedge {\tilde d}{\tilde\mcA}+\frac{2}{3}{\tilde\mcA}\wedge{\tilde\mcA}\wedge{\tilde\mcA}\right] \ ,
\ee
which then leads to the superpotential \cite{Witten:1985bz}
\be
                    W\sim \lambda \int_{CY} \Omega\wedge \tr_{E_8}\left[{\bar\mcA}\wedge {\bar \partial}{\bar\mcA}+\frac{2}{3}{\bar\mcA}\wedge{\bar\mcA}\wedge{\bar\mcA}\right] \ .
\ee
Note that \emph{a priori} this superpotential includes not only the bundle moduli and charged matter but also the massive modes. In addition, it also depends on the complex structure moduli. As pointed out in \cite{Witten:1985bz}, once the condition of holomorphicity ${\bar\mcF}=0$ is satisfied, the bundle moduli parameterize flat directions of the superpotential. In fact,
\be
             \partial_\phi W\sim 2\lambda\int \Omega\wedge\tr_{E_8} (\partial_\phi{\bar\mcA}\wedge{\bar\mcF}) \ ,
\ee
vanishes if ${\bar\mcF}(\phi)=0$.

It is instructive to see what kind of terms one obtains expanding around a given gauge bundle $\mcA(\phi)$,
\be
                   {\bar\mcA}={\bar\mcA}(\phi)+C\equiv{\bar\mcA}(\phi)+u_aC^a \ .
\ee
We find that
\be\label{eq:matter_super}
                W_{pert}\sim\lambda\int\Omega\wedge\tr_{E_8}\left(C\wedge {\bar D} C+\frac{2}{3}C\wedge C\wedge C\right) \ ,
\ee
already taking into account that the background gauge bundle is holomorphic. We recall that the charged massless fluctuations should satisfy ${\bar D} C=0$. Then, only the last term contributes to the superpotential which is a cubic function of the charged matter fields. It is important to stress that it also depends on the complex structure moduli $z$ and bundle moduli $\phi$.

In addition to \eqref{eq:matter_super}, the four-flux $G$ generally also induces a flux superpotential $W_{flux}(z)$ through \eqref{eq:flux_superpotential}, depending only on the complex structure moduli $z$. In this paper we will assume that all complex structure moduli are frozen by the effect of $W_{flux}$. After integrating out these moduli, a constant\footnote{See however ref.\cite{Anguelova:2006qf}, which indicates that the flux superpotential may also depend on the K\"ahler moduli.} superpotential $W_0$ remains, which can in 5d be seen as a Scherk-Schwarz twist.

\vspace{12pt}

The perturbative effects discussed sofar cannot alone provide for a stable vacuum. Fortunatly, a number of non-perturbative effects have the potential to deform the vacuum structure of heterotic M-theory. In particular, non-perturbative superpotentials can be induced by: i) euclidean M2 branes wrapping CY 2-cycles and stretching between the orbifold planes (and possible bulk M5 branes) \cite{Moore:2000fs,Lima:2001jc,Lima:2001nh,Curio:2001qi}; ii) gaugino condensation in strongly coupled gauge sector(s) \cite{Dine:1985rz,Kaplunovsky:1993rd,Brignole:1993dj}, typically at the hidden brane\cite{Horava:1996vs,Nilles:1997cm,Lalak:1997zu,Lukas:1997rb,Lukas:1999kt}; iii) and euclidean M5 branes wrapping the whole CY 3-fold at the orbifold fix points \cite{Buchbinder:2006xh}.

Regarding the third effect (M5 brane instantons), let us point out its similarity to the superpotential induced by the gauge instantons, in that both are controled by the $S_i$ moduli. One might wonder whether these are two different descriptions of the same phenomenon. For that reason, and since the detailed form of the euclidean M5 brane superpotential is not fully understood \cite{Buchbinder:2006xh}, we will neglect it and assume that the non-perturbative superpotential reads \cite{Moore:2000fs},\footnote{See, however, ref.\cite{Carlevaro:2005bk}, which advocates another form for the non-perturbative superpotential $W_{np}$.}
\be\label{eq:np_super}
                 W_{np}=\sum_i A_i e^{-aT^i}+Be^{-bS_{\pi}} \ ,
\ee
where the first term at the r.h.s. arises from the brane instantons while the second one is due to gaugino condensation at the hidden brane. Here, $T^i=\beta_I^i T^I$ is the K\"ahler modulus associated with the wrapped $i$-th holomorphic two cycle and $a$ is determined by the M2 brane tension.  The constant $b$ is determined by the field content at the hidden brane. On the other hand, the prefactors $A_i$ and $B$ are, in general, functions of additional moduli. That is
\be
                A_i=A_i(\phi,C,z),\quad B=B(\phi,C,z),
\ee
where $\phi,C$ denote brane localised bundle moduli and matter and $z$ stands for complex structure moduli. In case we have additional M5 branes in the 5d bulk, Eq.\eqref{eq:np_super} includes additional terms due to M2 brane instantons connecting the different branes.

The bundle moduli dependence of $A_i$ is in general not known but there are a few results that can be stated. Firstly, let us recall that for $A_i$ to be non-vanishing, the M2 brane instanton must be wrapping an isolated genus zero holomorphic curve (times the orbifold interval). Moreover, the vector bundle pulled back to the wrapped curve must be trivial which implies that since $T^i\equiv\beta^i_I T^I$, then \cite{Distler:1987ee}
\be
                    \beta_I^i q^I \sim \int c_1(V_X) \wedge \beta_I^i\,\omega^I = 0 \ ,
\ee
must be satisfied. This has the interesting implication that $A_i(\phi,C)$ must be a singlet.\footnote{We see that this presents a solution to the possible incompability between D-term potentials and non-perturbative superpotentials different from the one found in the case of gaugino condensation on D7-branes in type IIB compactifications. In this case, it was shown by an explicit calculation \cite{Haack:2006cy} that the prefactor can be charged under the anomalous U(1)$_X$ in such a way as to make the superpotential gauge invariant. In our case, the prefactor is allways a singlet and therefore must vanish in case the exponential factor in the superpotential is not U(1)$_X$ invariant.}

The prefactor $A_i$ is essentially determined by the Pfaffian of the chiral Dirac operator evaluated in the background of the gauge bundle pulled back to the holomorphic curve. In a series of papers \cite{Buchbinder:2002ji,Buchbinder:2002ic,Buchbinder:2002pr}, Buchbinder \emph{et al.} were able to explicitly calculate the bundle moduli dependence of $A_i$ in a number of examples with vector bundles constructed on elliptically fibered CY 3-folds. They found that $A_i$ is a non-vanishing homogeneous polynomial function of certain bundle moduli, dubbed \emph{transition} moduli since they trigger small instanton transitions at the $i$-th curve as they approach zero. A nice explanation of these and related issues can be found in \cite{Buchbinder:2003pi}.

Finally, let us point out that in this paper we assume the gauge bundle at the hidden brane to be trivial. Therefore, in this case the gaugino condensate 1-loop determinant $B$ is just a holomorphic function of the complex structure moduli.

\section{Supersymmetric (adS) vacua}\label{sec:susy_vacua}

In this section we search for supersymmetric vacua. In view of the new understanding of the K\"ahler potential with brane multiplets obtained in section \ref{subsec:kaehler_brane}, we expect our results to be distinct from the ones in ref.\cite{Buchbinder:2003pi}. To keep our dicussion as simple as possible we take $h^{1,1}=1$ and no bulk M5 branes. Furthermore, we consider the gauge bundle at the hidden brane to be trivial and exclude the possibility of \emph{anomalous} U(1) factors acting at the visible brane. We follow \cite{Buchbinder:2003pi}, which in the spirit of KKLT \cite{Kachru:2003aw} assumes that the complex structure moduli are stabilized in values which are approximately independent of the other moduli. Then, the superpotential reads
\be
               W=W_0+W_{pert}(\phi,C)+A(\phi,C)e^{-aT}+Be^{-bS_\pi}.
\ee
Supersymmetric anti-de Sitter vacua can be obtained whenever
\be
            D_i W=0
\ee
can be satisfied, with $i$ running over all moduli.

We observe first that $D_{C^a}W=0$ can be satisfied at $C=0$ in case $A(\phi,C)$ is regular at $C=0$, which we take for granted in the following. In fact the latter implies that $\partial_{C^a}W=0$. Then, since (cf. \eqref{eq:kaehler_metric})
\be
                 \mcK_a\equiv\partial_{C^a}\mcK=-\mcK_IG^I_{a\bar{b}}(\phi,\bar{\phi})\bar{C}^{\bar{b}}
\ee
also vanishes for $C=0$, we find that $\mcK_a W=0$ at that point and $D_{C^a}W=0$ as we wanted to show.

Our search for supersymmetric (AdS) vacua reduces now to studying the superpotential
\be
              W={W}_0+A(\phi)e^{-aT}+Be^{-bS_\pi} \ ,
\ee
where $A(\phi)=A(\phi,C=0)$. Recall that in section \ref{subsec:superpotential} we pointed out that $W_{pert}(\phi,C=0)$ would not depend on the vector bundle moduli $\phi$ once we impose the vector bundle to be holomorphic (${\bar\mcF}=0$). We will assume this to be the case in the following. It is then not difficult to find that, due to the specific way the bundle moduli and the K\"ahler moduli mix in Eq.\eqref{eq:kaehler_general_brane_stuff}, we have
\be
              D_{\phi^\alpha}W=e^{-aT}\left(\partial_\alpha-a k_\alpha\right)A(\phi)-k_\alpha D_TW \ .
\ee
The value of $\phi^\alpha$ in a supersymmetric vacuum (with $D_TW=0$) is thus found by solving the \emph{reduced} problem \footnote{The analog result, in the context of type IIB supersymmetric vacua with D3-branes, was obtained in \cite{DeWolfe:2007hd}.}
\be\label{eq:reduced_problem}
              \tilde{D}_\alpha \mcW(\phi)\equiv \partial_\alpha\mcW(\phi)+ k_\alpha\mcW(\phi)=0 \ ,
\ee
with the reduced superpotential $\mcW(\phi)=A(\phi)^{-\frac{1}{a}}$ and the reduced K\"ahler potential being $k(\phi,\bar{\phi})$. It is important to stress that the value of $\phi^\alpha$ in the supersymmetric vacuum is fully independent of the geometric moduli $S_\pi$ and $T$. This is in contrast to the findings of \cite{Buchbinder:2003pi}.

Moreover, the above supersymmetry condition on the bundle moduli has an interesting physical interpretation, as we explain now. This relies on understanding the origin of the $\phi$-dependence of the prefactor $A(\phi)$. Adapting an argument given in \cite{Giddings:2005ff,Baumann:2006th}, in the type IIB context, we find that this dependence can be traced back to the correction that the volume of the 3-cycle wrapped by the brane instanton suffers in the presence of the non-trivial brane localised gauge bundle\footnote{In the type IIB case \cite{Giddings:2005ff,Baumann:2006th}, it is the volume of wrapped 4-cycles which is deformed in the presence of a D3-brane.}. In the present case, with $h^{1,1}=1$, the total volume reads
\be
             V_3\equiv \oint dy V_{1,1}= V_3^0+\delta V_3^0 \ ,
\ee
with the \emph{undeformed} volume being (cf. \eqref{eq:undeformed_cycle})
\be
              V_3^0=T+\bar{T}-k(\phi,\bar{\phi}) \ ,
\ee
and the correction
\be
           \delta V_3^0=k(\phi,\bar{\phi})-a^{-1}\ln|A(\phi)|^2+const.
\ee
Clearly, the supersymmetry condition on the gauge bundle, $\tilde{D}_\alpha \mcW(\phi)=0$, now gains a suggestive new face
\be
               \frac{\partial}{\partial\phi^\alpha}\,\delta V_3^0(\phi,\bar{\phi})=0 \ .
\ee
We thus see that in a supersymmetric vacuum, the bundle moduli extremize the volume of the cycles wrapped by brane instantons.

In case a solution, $\phi=\phi_0$ to Eq.\eqref{eq:reduced_problem} can be found, we are left with a set of two (complex) coupled equations involving just the moduli $S_\pi$ and $T$. For simplicity, in the following we work with the moduli $S_\pi$ and $S_0=S_\pi+\alpha T$. The conditions $D_iW=0$ for a supersymmetric vacuum can be combined to give
\be
              S_0=\left(1+\frac{\alpha b}{a}\right)S_\pi+\frac{\alpha}{a}\ln\left\{\frac{aA_0}{\alpha bB}\left(1-\left(\frac{S_\pi+{\bar S}_\pi}{S_0+{\bar S}_0}\right)^{\frac{1}{3}}\right)\right\} \ ,
\ee
where $A_0\equiv A(\phi_0)$, and
\be
              {W}_0=-\left[ 1+\frac{a}{\alpha b} + \frac{a}{4\alpha} (S_0+\bar{S}_0)-\left(\frac{a}{\alpha b} + \frac{a}{4\alpha} (S_\pi+\bar{S}_\pi) \right)\left(\frac{S_\pi+{\bar S}_\pi}{S_0+{\bar S}_0}\right)^{\frac{1}{3}}\right]A_0\, e^{-\frac{a}{\alpha}(S_0-S_\pi)} \ .
\ee
Following \cite{Buchbinder:2003pi}, we have $a\sim 10^2$, $b\sim 5$ and $B\sim 10^{-6}M_p^3$. On the other hand, acceptable phenomenology imposes that the brane tension $\alpha$ is bounded as $\alpha\lesssim 1$ and $\textup{Re}(S_0)\sim 1$. With these values we find that there is an AdS$_4$ supersymmetric vacuum with
\be\label{eq:S_pi}
             \textup{Re}(S_\pi)\simeq\left(1-\frac{\alpha b}{a}\right)\textup{Re}(S_0)\approx \textup{Re}(S_0) \ ,
\ee
if we fine tune the flux superpotential $W_0$ to be
\be
             |W_0|\sim|B|e^{-b\textup{Re}(S_\pi)}\sim 10^{-8}M_p^3 \ .
\ee
We note, however, that the value of $\textup{Re}(S_\pi)$ is by far too large to be phenomenologically acceptable. In fact, the problem lies on the ratio $(b/a)\sim 5\cdot 10^{-2}$ being too small. That is, the force exerted by the membrane instanton against the growth of the wrapped cycle is too strong when compared to the effect of gaugino condensation. The situation would improve if we manage to bring $b/a$ to be of order one. It is  conceivable to increase $b$ by breaking down the $E_8$ at the hidden brane to some subgroup by using a non-trivial gauge bundle. To decrease $a$ by the needed amount, one would have to decrease the size of the holomorphic curve wrapped by the brane instanton by one order of magnitude. It is interesting to note that the vacuum \eqref{eq:S_pi} depends only very weakly on $A_0=A(\phi_0)$ and therefore, in principle, we cannot choose the visible sector gauge bundle in such a way as to \emph{move} the susy vacuum into a phenomenologically interesting corner of the moduli space.

Finally, it is straightforward to determine the value of the potential energy at the vacuum. We find
\be
                V_{min}\sim -\left(\frac{a}{b}\right)^3\frac{|W_0|^2}{M_p^2}\sim - 10^{-12} M_p^4 \ ,
\ee
a rather deep $AdS$ vacuum.

Before closing this section let us comment on the assumptions we made in obtaining the above results. The first assumption was that the complex structure moduli can be fixed due to the effect of a flux superpotential with values that do not depend on the other moduli $\phi,T,S_\pi$. Clearly, if this was not to be the case, the above results could be strongly modified. We refer the reader to \cite{Choi:2004sx} for a nice discussion of this possibility. Secondly, we assumed that the charged scalars can be stabilized at $C=0$. This relies on the fact that $A(z,\phi,C)$ is a gauge singlet and smooth at $C=0$ and on the assumption that the gauge bundle at the hidden brane is \emph{trivial}. Non trivial bundles may allow for more exotic possibilities, with a charged matter dependent prefactor $B$, obstructing supersymmetric vacua. The third assumption was to exclude the possibility that Eq.\eqref{eq:reduced_problem} has no solution for holomorphic vector bundles satisfying (the zero-slope limit of) the hermitian Yang-Mills equations. In fact, the possibility that non-perturbative corrections might destabilize heterotic vacua was raised long ago \cite{Ellis:1986pv,Dine:1986zy,Dine:1987bq} (see also \cite{Distler:1986wm,Distler:1987ee}). However, we should note that \cite{Buchbinder:2003pi} gave reasonable arguments against that possibility at least for the class of compactifications studied therein.

\section{Discussion}

In this paper we performed a discussion of the ingredients involved in moduli stabilization in heterotic M-theory. That is, we derived expressions for the low-energy K\"ahler potential and superpotentials as functions of the dilaton $S$ and K\"ahler moduli $T^I$, as well as of the bundle moduli $\phi$ and charged matter $C$. For this purpose we used a superfield description of the relevant couplings of 5d heterotic M-theory, which in this paper we expanded beyond the standard embedding.

One of the main achievements of this paper is the first time derivation of the way the bundle moduli and charged matter enter the K\"ahler potential. We found in section \ref{sec:brane_localised} that the K\"ahler potential of heterotic M-theory (and its weak coupling limit, the heterotic string) is similar in form to the De Wolfe-Giddings K\"ahler potential \cite{DeWolfe:2002nn} in type IIB string, but unlike the latter it is defined also for $h^{1,1}>0$. Using this and the perturbative and non-perturbative superpotentials discussed in section \ref{subsec:superpotential} we performed a study of moduli stabilization, searching for supersymmetric vacua (in a setup with $h^{1,1}=1$ for simplicity). Let us recapitulate our main findings.

The configuration of the gauge bundle is determined by solving a "reduced" supersymmetry condition, which involves only the bundle moduli $\phi$. We have argued that this condition has a simple physical interpretation. Namely, it picks extrema of the volume of the isolated holomorphic curves (wrapped by membrane instantons) in respect to the bundle moduli. If a solution ${\bar\mcA}(\phi_0)$ is found, we can determine the values of the geometric moduli $S$ and $T^I$ in the vacuum, by solving the corresponding susy conditions in the ${\bar\mcA}(\phi_0)$ background. We found that, at least in the $h^{1,1}=1$ case without five-branes in the bulk, at the supersymmetric AdS$_4$ vacuum the geometric moduli take values far from the phenomenologically acceptable corner of the moduli space.

Anomalous U(1)'s are predicted in recent semi-realistic heterotic model-building. In this paper we derived the FI D-terms induced by the anomalous U(1)'s, extending to strong coupling the results of \cite{Blumenhagen:2005ga}. As we pointed out in section \ref{sec:FI}, the D-term potential drives the theory towards strong coupling and may therefore be of use in moduli stabilization. We have also seen that the appearence of a charged K\"ahler modulus $T^i$ implies the vanishing of the membrane instanton superpotential associated with $T^i$. We note that, interestingly, in that case a contracting force (the membrane instanton) is traded for an expanding one (the FI D-term).

To obtain realistic vacua, the AdS supersymmetric minima discussed in this paper have not only to be uplifted to de Sitter vacua, but also the Vev's of the geometric moduli must be changed. Several ways of achieving this have already been proposed in the literature, mostly in the realm of type IIB compactifications. These include: D-terms from anomalous U(1)'s \cite{Buchbinder:2004im} (\cite{Burgess:2003ic,Dudas:2006vc,Achucarro:2006zf} in the type IIB context); anti M5-branes \cite{Braun:2006th,Gray:2007mg} (${\bar D3}$ in KKLT \cite{Kachru:2003aw}); de Sitter vacua from matter superpotentials \cite{Lebedev:2006qq}; and loop corrections \cite{Anguelova:2005jr} (or the combination of $\alpha'$ and loop corrections \cite{Balasubramanian:2004uy,Bobkov:2004cy,Balasubramanian:2005zx,vonGersdorff:2005bf,Berg:2005yu,Parameswaran:2006jh,Dudas:2006vc,Westphal:2006tn} in type IIB compactifications). We will investigate some of these and other possibilities in a forthcoming publication, where we will address the construction of stable de Sitter vacua in 5d heterotic M-theory (see \cite{Correia:2006vf} for an explicit example of dS stable vacua in 5d gauged supergravity at one loop).

Since to obtain de Sitter vacua we will have to find supersymmetry breaking minima of the scalar potential, the question arises as to whether the gauge bundle is still determined \emph{independently} of the geometric moduli, as in the AdS vacua studied in section \ref{sec:susy_vacua}. Remarkably, it can be shown (when $C^a=0$) that if the scalar potential has an extremum in respect to $S_\pi$ and $T$, i.e. $\partial_{S_\pi} V=\partial_T V=0$, then the reduced flatness condition, Eq.\eqref{eq:reduced_problem}, will imply that $V$ also has an extremum in respect to the bundle moduli, $\partial_\alpha V=0$, and we expect this to be the case also for $h^{1,1}>1$. We thus see that the gauge bundle in a de Sitter vacuum with $C=0$ is exactly the same as in the supersymmetric AdS vacuum, which is determined by the reduced flatness condition. However, it remains to be seen what happens once we allow for vacua where the charged scalars receive non-vanishing Vev's as in \cite{Becker:2004gw}, or include anti M5 branes in the 5d bulk \cite{Braun:2006th,Gray:2007mg}. In this case we cannot exclude that the configuration of the gauge bundle will be dependent on the geometric moduli.

\section*{Acknowledgements}
This work was started while F.P.C. was visiting the Institut f\"ur Theoretische Physik, Heidelberg, Germany, with the support of Schwerpunktprogramm "Stringtheorie ..." (5PP1096). We are grateful to M. Olechowski and Z. Tavartkiladze for interesting discussions. F.P.C is supported by FCT through the grant SFRH/BPD/20667/2004.

\appendix

\section{The structure of the K\"ahler potential}\label{app:A}

It is well-known that in the type IIB compactifications the no-scale structure is preserved even in the presence of a mobile D3 brane. Here we prove the heterotic M-theory counterpart of this result, namely that
\be
               \mcK^{i{\bar j}}\mcK_i\mcK_{\bar j}=4,
\ee
where $i,\bar{j}$ run over all moduli, even in the presence of the brane fields. Our sole assumption is that in the absence of brane fields we have $\mcK_i(X^i+\bar{X}^i)=-n$, where $X^i=T^I,S_0$. In case of interest we have $n=4$, but our analysis also includes the type IIB case with $n=3$. We will now prove this assumption in the 5d heterotic M-theory case. The crucial fact is that
\be
               \mcK=-3\ln\Omega \ ,
\ee
where
\be\label{eq:Omega}
               \Omega=\Omega_0\cdot F\left(\alpha_0\frac{T^I+{\bar T}^I}{S_0+{\bar S}_0}\right) \ ,
\ee
and $\Omega_0$ is the well known weak coupling result
\be
               \Omega_0^3=(S_0+{\bar S}_0)\,d_{IJK}(T^I+{\bar T}^I)(T^J+{\bar T}^
               J)(T^K+{\bar T}^K) \ .
\ee
In other words, $F(0)=1$. Note that, without any loss of generality, we chose $\alpha_I=0$ for $I\neq 0$. Using Eq.\eqref{eq:Omega}, it is then straightforward to find that
\be
                \mcK_i(X^i+\bar{X}^i)=-4 \ .
\ee

As we will show now, once we include brane fields the same result follows due to the following structure of the K\"ahler potential
\be
               \mcK=\mcK\left(S_0+{\bar S}_0-\alpha_Ik^I(\phi,\bar\phi ; C,\bar{C})\,;T^I+{\bar T}^I-k^I(\phi,\bar\phi ; C,\bar{C}) \right).
\ee
A word on the notation: We denote $X^I\equiv T^I$ and $X^{I=-1}\equiv S_0$, while $i$ runs over all chiral superfields. In addition, $k^{I=-1}\equiv\alpha_I k^I$. We have thus that
\be\label{eq:startingpoint}
               \mcK_I(X^I+\bar{X}^I-k^I)=-n.
\ee
It follows then that
\be
               \mcK_{I\bar{j}}(X^I+\bar{X}^I-k^I)=-\mcK_{\bar{j}},
\ee
and therefore
\be
               \mcK^{i\bar{j}}\mcK_{\bar{j}}=-\delta^i_I(X^I+\bar{X}^I-k^I).
\ee
Using Eq.\eqref{eq:startingpoint} we finally find that
\be
                      \mcK^{i\bar{j}}\mcK_i\mcK_{\bar{j}}=n,
\ee
as we wanted to show.


\begin{thebibliography}{99}

\bibitem{Kachru:2003aw}
  S.~Kachru, R.~Kallosh, A.~Linde and S.~P.~Trivedi,
  \emph{De Sitter vacua in string theory},
  Phys.\ Rev.\  D {\bf 68} (2003) 046005
  [hep-th/0301240].

\bibitem{Horava:1995qa}
  P.~Horava and E.~Witten,
  \emph{Heterotic and type I string dynamics from eleven dimensions},
  Nucl.\ Phys.\  B {\bf 460} (1996) 506
  [hep-th/9510209].

\bibitem{Horava:1996ma}
  P.~Horava and E.~Witten,
  \emph{Eleven-Dimensional Supergravity on a Manifold with Boundary},
  Nucl.\ Phys.\  B {\bf 475} (1996) 94
  [hep-th/9603142].

\bibitem{Witten:1996mz}
  E.~Witten,
  \emph{Strong Coupling Expansion Of Calabi-Yau Compactification},
  Nucl.\ Phys.\  B {\bf 471} (1996) 135
  [hep-th/9602070].

\bibitem{Banks:1996ss}
  T.~Banks and M.~Dine,
  \emph{Couplings and Scales in Strongly Coupled Heterotic String Theory},
  Nucl.\ Phys.\  B {\bf 479} (1996) 173
  [hep-th/9605136].

\bibitem{Lukas:1998yy}
  A.~Lukas, B.~A.~Ovrut, K.~S.~Stelle and D.~Waldram,
  \emph{The universe as a domain wall},
  Phys.\ Rev.\  D {\bf 59} (1999) 086001
  [hep-th/9803235].

\bibitem{Lukas:1998tt}
  A.~Lukas, B.~A.~Ovrut, K.~S.~Stelle and D.~Waldram,
  \emph{Heterotic M-theory in five dimensions},
  Nucl.\ Phys.\  B {\bf 552} (1999) 246
  [hep-th/9806051].

\bibitem{Giddings:2001yu}
  S.~B.~Giddings, S.~Kachru and J.~Polchinski,
  \emph{Hierarchies from fluxes in string compactifications},
  Phys.\ Rev.\  D {\bf 66} (2002) 106006
  [hep-th/0105097].

\bibitem{Burgess:2003ic}
  C.~P.~Burgess, R.~Kallosh and F.~Quevedo,
  \emph{de Sitter string vacua from supersymmetric D-terms},
  JHEP {\bf 0310} (2003) 056
  [hep-th/0309187].
    
\bibitem{Saltman:2004sn}
  A.~Saltman and E.~Silverstein,
  \emph{The scaling of the no-scale potential and de Sitter model building},
  JHEP {\bf 0411} (2004) 066
  [hep-th/0402135].
  
\bibitem{Balasubramanian:2004uy}
  V.~Balasubramanian and P.~Berglund,
  \emph{Stringy corrections to Kaehler potentials, SUSY breaking, and the cosmological constant problem},
  JHEP {\bf 0411} (2004) 085
  [hep-th/0408054].

\bibitem{Choi:2004sx}
  K.~Choi, A.~Falkowski, H.~P.~Nilles, M.~Olechowski and S.~Pokorski,
  \emph{Stability of flux compactifications and the pattern of supersymmetry breaking},
  JHEP {\bf 0411} (2004) 076
  [hep-th/0411066].
  
\bibitem{Bobkov:2004cy}
  K.~Bobkov,
  \emph{Volume stabilization via alpha' corrections in type IIB theory with fluxes},
  JHEP {\bf 0505} (2005) 010
  [hep-th/0412239].

\bibitem{Balasubramanian:2005zx}
  V.~Balasubramanian, P.~Berglund, J.~P.~Conlon and F.~Quevedo,
  \emph{Systematics of moduli stabilisation in Calabi-Yau flux compactifications},
  JHEP {\bf 0503} (2005) 007
  [hep-th/0502058].

\bibitem{vonGersdorff:2005bf}
  G.~von Gersdorff and A.~Hebecker,
  \emph{Kaehler corrections for the volume modulus of flux compactifications},
  Phys.\ Lett.\  B {\bf 624} (2005) 270
  [hep-th/0507131].

\bibitem{Berg:2005yu}
  M.~Berg, M.~Haack and B.~Kors,
  \emph{On volume stabilization by quantum corrections},
  Phys.\ Rev.\ Lett.\  {\bf 96} (2006) 021601
  [hep-th/0508171].

\bibitem{Achucarro:2006zf}
  A.~Achucarro, B.~de Carlos, J.~A.~Casas and L.~Doplicher,
  \emph{de Sitter vacua from uplifting D-terms in effective supergravities from
  realistic strings},
  JHEP {\bf 0606} (2006) 014
  [hep-th/0601190].

\bibitem{Parameswaran:2006jh}
  S.~L.~Parameswaran and A.~Westphal,
  \emph{de Sitter string vacua from perturbative Kaehler corrections and consistent D-terms},
  JHEP {\bf 0610} (2006) 079
  [hep-th/0602253].

\bibitem{Lebedev:2006qq}
  O.~Lebedev, H.~P.~Nilles and M.~Ratz,
  \emph{de Sitter vacua from matter superpotentials},
  Phys.\ Lett.\  B {\bf 636} (2006) 126
  [hep-th/0603047].

\bibitem{Dudas:2006vc}
  E.~Dudas and Y.~Mambrini,
  \emph{Moduli stabilization with positive vacuum energy},
  JHEP {\bf 0610} (2006) 044
  [hep-th/0607077].

\bibitem{Westphal:2006tn}
  A.~Westphal,
  \emph{de Sitter String Vacua from Kahler Uplifting},
  JHEP {\bf 0703} (2007) 102
  [hep-th/0611332].

\bibitem{Curio:2001qi}
  G.~Curio and A.~Krause,
  \emph{G-fluxes and non-perturbative stabilisation of heterotic M-theory},
  Nucl.\ Phys.\  B {\bf 643} (2002) 131
  [hep-th/0108220].

\bibitem{Curio:2003ur}
  G.~Curio and A.~Krause,
  \emph{Enlarging the parameter space of heterotic M-theory flux compactifications to phenomenological viability},
  Nucl.\ Phys.\  B {\bf 693} (2004) 195
  [hep-th/0308202].

\bibitem{Buchbinder:2003pi}
  E.~I.~Buchbinder and B.~A.~Ovrut,
  \emph{Vacuum stability in heterotic M-theory},
  Phys.\ Rev.\  D {\bf 69} (2004) 086010
  [hep-th/0310112].

\bibitem{Becker:2004gw}
  M.~Becker, G.~Curio and A.~Krause,
  \emph{De Sitter vacua from heterotic M-theory},
  Nucl.\ Phys.\  B {\bf 693} (2004) 223
  [hep-th/0403027].

\bibitem{Buchbinder:2004im}
  E.~I.~Buchbinder,
  \emph{Raising anti de Sitter vacua to de Sitter vacua in heterotic M-theory},
  Phys.\ Rev.\  D {\bf 70} (2004) 066008
  [hep-th/0406101].

\bibitem{Anguelova:2005jr}
  L.~Anguelova and D.~Vaman,
  \emph {R**4 corrections to heterotic M-theory},
  Nucl.\ Phys.\  B {\bf 733} (2006) 132
  [hep-th/0506191].

\bibitem{Braun:2006th}
  V.~Braun and B.~A.~Ovrut,
  \emph{Stabilizing moduli with a positive cosmological constant in heterotic
  M-theory},
  JHEP {\bf 0607} (2006) 035
  [hep-th/0603088].

\bibitem{Curio:2006dc}
  G.~Curio and A.~Krause,
  \emph{S-Track stabilization of heterotic de Sitter vacua},
  hep-th/0606243.

\bibitem{DeWolfe:2002nn}
  O.~DeWolfe and S.~B.~Giddings,
  \emph{Scales and hierarchies in warped compactifications and brane worlds},
  Phys.\ Rev.\  D {\bf 67} (2003) 066008
  [hep-th/0208123].

\bibitem{Blumenhagen:2005ga}
  R.~Blumenhagen, G.~Honecker and T.~Weigand,
  \emph{Loop-corrected compactifications of the heterotic string with line
  bundles},
  JHEP {\bf 0506} (2005) 020
  [hep-th/0504232].

\bibitem{Blumenhagen:2006ux}
  R.~Blumenhagen, S.~Moster and T.~Weigand,
  \emph{Heterotic GUT and standard model vacua from simply connected Calabi-Yau
  manifolds},
  Nucl.\ Phys.\  B {\bf 751} (2006) 186
  [hep-th/0603015].

\bibitem{Tatar:2006dc}
  R.~Tatar and T.~Watari,
  \emph{Proton decay, Yukawa couplings and underlying gauge symmetry in string
  theory},
  Nucl.\ Phys.\  B {\bf 747} (2006) 212
  [hep-th/0602238].

\bibitem{Correia:2006pj}
  F.~Paccetti~Correia, M.~G.~Schmidt and Z.~Tavartkiladze,
  \emph{4D superfield reduction of 5D orbifold SUGRA and heterotic M-theory},
  Nucl.\ Phys.\  B {\bf 751} (2006) 222
  [hep-th/0602173].

\bibitem{PaccettiCorreia:2004ri}
  F.~Paccetti Correia, M.~G.~Schmidt and Z.~Tavartkiladze,
  \emph{Superfield approach to 5D conformal SUGRA and the radion},
  Nucl.\ Phys.\  B {\bf 709} (2005) 141
  [hep-th/0408138].

\bibitem{Correia:2004pz}
  F.~Paccetti~Correia, M.~G.~Schmidt and Z.~Tavartkiladze,
  \emph{(BPS) Fayet-Iliopoulos terms in 5D orbifold SUGRA},
  Phys.\ Lett.\  B {\bf 613} (2005) 83
  [hep-th/0410281].

\bibitem{Abe:2004ar}
  H.~Abe and Y.~Sakamura,
  \emph{Superfield description of 5D supergravity on general warped geometry},
  JHEP {\bf 0410} (2004) 013
  [hep-th/0408224].

\bibitem{Abe:2006eg}
  H.~Abe and Y.~Sakamura,
  \emph{Roles of Z(2)-odd N = 1 multiplets in off-shell dimensional reduction of 5D supergravity},
  Phys.\ Rev.\  D {\bf 75} (2007) 025018
  [hep-th/0610234].

\bibitem{Kugo:2000hn}
  T.~Kugo and K.~Ohashi,
  \emph{Supergravity tensor calculus in 5D from 6D},
  Prog.\ Theor.\ Phys.\  {\bf 104} (2000) 835
  [hep-ph/0006231].

\bibitem{Kugo:2000af}
  T.~Kugo and K.~Ohashi,
  \emph{Off-shell d = 5 supergravity coupled to matter-Yang-Mills system},
  Prog.\ Theor.\ Phys.\  {\bf 105} (2001) 323
  [hep-ph/0010288].

\bibitem{Fujita:2001bd}
  T.~Fujita, T.~Kugo and K.~Ohashi,
  \emph{Off-shell formulation of supergravity on orbifold},
  Prog.\ Theor.\ Phys.\  {\bf 106} (2001) 671
  [hep-th/0106051].

\bibitem{Kugo:2002js}
  T.~Kugo and K.~Ohashi,
  \emph{Superconformal tensor calculus on orbifold in 5D},
  Prog.\ Theor.\ Phys.\  {\bf 108} (2002) 203
  [hep-th/0203276].

\bibitem{Bagger:2006hm}
  J.~Bagger and C.~Xiong,
  \emph{N = 2 nonlinear sigma models in N = 1 superspace: Four and five
  dimensions},
  hep-th/0601165.

\bibitem{Lalak:2001dv}
  Z.~Lalak,
  \emph{Low energy supersymmetry in warped brane worlds},
  hep-th/0109074.

\bibitem{Lukas:1998hk}
  A.~Lukas, B.~A.~Ovrut and D.~Waldram,
  \emph{Non-standard embedding and five-branes in heterotic M-theory},
  Phys.\ Rev.\  D {\bf 59} (1999) 106005
  [hep-th/9808101].

\bibitem{Donagi:1998xe}
  R.~Donagi, A.~Lukas, B.~A.~Ovrut and D.~Waldram,
  \emph{Non-perturbative vacua and particle physics in M-theory},
  JHEP {\bf 9905} (1999) 018
  [hep-th/9811168].

\bibitem{Donagi:1999gc}
  R.~Donagi, A.~Lukas, B.~A.~Ovrut and D.~Waldram,
  \emph{Holomorphic vector bundles and non-perturbative vacua in M-theory},
  JHEP {\bf 9906} (1999) 034
  [hep-th/9901009].

\bibitem{Donagi:2000zf}
  R.~Donagi, B.~A.~Ovrut, T.~Pantev and D.~Waldram,
  \emph{Standard-model bundles on non-simply connected Calabi-Yau threefolds},
  JHEP {\bf 0108} (2001) 053
  [hep-th/0008008].

\bibitem{Derendinger:2000gy}
  J.~P.~Derendinger and R.~Sauser,
  \emph{A five-brane modulus in the effective N = 1 supergravity of M-theory},
  Nucl.\ Phys.\  B {\bf 598} (2001) 87
  [hep-th/0009054].

\bibitem{Brandle:2001ts}
  M.~Brandle and A.~Lukas,
  \emph{Five-branes in heterotic brane-world theories},
  Phys.\ Rev.\  D {\bf 65} (2002) 064024
  [hep-th/0109173].

\bibitem{Moore:2000fs}
  G.~W.~Moore, G.~Peradze and N.~Saulina,
  \emph{Instabilities in heterotic M-theory induced by open membrane instantons},
  Nucl.\ Phys.\  B {\bf 607} (2001) 117
  [hep-th/0012104].

\bibitem{Kugo:1982mr}
  T.~Kugo and S.~Uehara,
  \emph{Improved Superconformal Gauge Conditions In The N=1 Supergravity Yang-Mills Matter System},
  Nucl.\ Phys.\  B {\bf 222} (1983) 125.

\bibitem{Gray:2003vw}
  J.~Gray and A.~Lukas,
  \emph{Gauge five brane moduli in four-dimensional heterotic models},
  Phys.\ Rev.\  D {\bf 70} (2004) 086003
  [hep-th/0309096].

\bibitem{Lukas:1999nh}
  A.~Lukas and K.~S.~Stelle,
  \emph{Heterotic anomaly cancellation in five dimensions},
  JHEP {\bf 0001} (2000) 010
  [hep-th/9911156].

\bibitem{Gukov:1999ya}
  S.~Gukov, C.~Vafa and E.~Witten,
  \emph{CFT's from Calabi-Yau four-folds},
  Nucl.\ Phys.\  B {\bf 584} (2000) 69
  [Erratum-ibid.\  B {\bf 608} (2001) 477]
  [hep-th/9906070].

\bibitem{Behrndt:2000zh}
  K.~Behrndt and S.~Gukov,
  \emph{Domain walls and superpotentials from M theory on Calabi-Yau three-folds},
  Nucl.\ Phys.\  B {\bf 580} (2000) 225
  [hep-th/0001082].

\bibitem{Becker:2002jj}
  M.~Becker and D.~Constantin,
  \emph{A note on flux induced superpotentials in string theory},
  JHEP {\bf 0308} (2003) 015
  [hep-th/0210131].

\bibitem{Anguelova:2006qf}
  L.~Anguelova and K.~Zoubos,
  \emph{Flux superpotential in heterotic M-theory},
  Phys.\ Rev.\  D {\bf 74} (2006) 026005
  [hep-th/0602039].

\bibitem{Witten:1985bz}
  E.~Witten,
  \emph{New Issues In Manifolds Of SU(3) Holonomy},
  Nucl.\ Phys.\  B {\bf 268} (1986) 79.

\bibitem{Lima:2001jc}
  E.~Lima, B.~A.~Ovrut, J.~Park and R.~Reinbacher,
  \emph{Non-perturbative superpotential from membrane instantons in heterotic M-theory},
  Nucl.\ Phys.\  B {\bf 614} (2001) 117
  [hep-th/0101049].

\bibitem{Lima:2001nh}
  E.~Lima, B.~A.~Ovrut and J.~Park,
  \emph{Five-brane superpotentials in heterotic M-theory},
  Nucl.\ Phys.\  B {\bf 626} (2002) 113
  [hep-th/0102046].

\bibitem{Dine:1985rz}
  M.~Dine, R.~Rohm, N.~Seiberg and E.~Witten,
  \emph{Gluino Condensation In Superstring Models},
  Phys.\ Lett.\  B {\bf 156} (1985) 55.

\bibitem{Kaplunovsky:1993rd}
  V.~S.~Kaplunovsky and J.~Louis,
  \emph{Model independent analysis of soft terms in effective supergravity and in string theory},
  Phys.\ Lett.\  B {\bf 306} (1993) 269
  [hep-th/9303040].

\bibitem{Brignole:1993dj}
  A.~Brignole, L.~E.~Ibanez and C.~Munoz,
  \emph{Towards a theory of soft terms for the supersymmetric Standard Model},
  Nucl.\ Phys.\  B {\bf 422} (1994) 125
  [Erratum-ibid.\  B {\bf 436} (1995) 747]
  [hep-ph/9308271].

\bibitem{Horava:1996vs}
  P.~Horava,
  \emph{Gluino condensation in strongly coupled heterotic string theory},
  Phys.\ Rev.\  D {\bf 54} (1996) 7561
  [hep-th/9608019].

\bibitem{Nilles:1997cm}
  H.~P.~Nilles, M.~Olechowski and M.~Yamaguchi,
  \emph{Supersymmetry breaking and soft terms in M-theory},
  Phys.\ Lett.\  B {\bf 415} (1997) 24
  [hep-th/9707143].

\bibitem{Lalak:1997zu}
  Z.~Lalak and S.~Thomas,
  \emph{Gaugino condensation, moduli potential and supersymmetry breaking in M-theory models},
  Nucl.\ Phys.\  B {\bf 515} (1998) 55
  [hep-th/9707223].

\bibitem{Lukas:1997rb}
  A.~Lukas, B.~A.~Ovrut and D.~Waldram,
  \emph{Gaugino condensation in M-theory on S**1/Z(2)},
  Phys.\ Rev.\  D {\bf 57} (1998) 7529
  [hep-th/9711197].

\bibitem{Lukas:1999kt}
  A.~Lukas, B.~A.~Ovrut and D.~Waldram,
  \emph{Five-branes and supersymmetry breaking in M-theory},
  JHEP {\bf 9904} (1999) 009
  [hep-th/9901017].

\bibitem{Buchbinder:2006xh}
  E.~I.~Buchbinder,
  \emph{Derivative F-terms from heterotic M-theory five-brane instanton},
  Phys.\ Lett.\  B {\bf 645} (2007) 281
  [hep-th/0611119].

\bibitem{Carlevaro:2005bk}
  L.~Carlevaro and J.~P.~Derendinger,
  \emph{Five-brane thresholds and membrane instantons in four-dimensional heterotic M-theory},
  Nucl.\ Phys.\  B {\bf 736} (2006) 1
  [hep-th/0502225].

\bibitem{Distler:1987ee}
  J.~Distler and B.~R.~Greene,
  \emph{Aspects of (2,0) String Compactifications},
  Nucl.\ Phys.\  B {\bf 304} (1988) 1.

\bibitem{Haack:2006cy}
  M.~Haack, D.~Krefl, D.~Lust, A.~Van Proeyen and M.~Zagermann,
  \emph{Gaugino condensates and D-terms from D7-branes},
  JHEP {\bf 0701} (2007) 078
  [hep-th/0609211].

\bibitem{Buchbinder:2002ji}
  E.~Buchbinder, R.~Donagi and B.~A.~Ovrut,
  \emph{Vector bundle moduli and small instanton transitions},
  JHEP {\bf 0206} (2002) 054
  [hep-th/0202084].

\bibitem{Buchbinder:2002ic}
  E.~I.~Buchbinder, R.~Donagi and B.~A.~Ovrut,
  \emph{Superpotentials for vector bundle moduli},
  Nucl.\ Phys.\  B {\bf 653} (2003) 400
  [hep-th/0205190].

\bibitem{Buchbinder:2002pr}
  E.~I.~Buchbinder, R.~Donagi and B.~A.~Ovrut,
  \emph{Vector bundle moduli superpotentials in heterotic superstrings and M-theory},
  JHEP {\bf 0207} (2002) 066
  [hep-th/0206203].

\bibitem{DeWolfe:2007hd}
  O.~DeWolfe, L.~McAllister, G.~Shiu and B.~Underwood,
  \emph{D3-brane Vacua in Stabilized Compactifications},
  hep-th/0703088.

\bibitem{Giddings:2005ff}
  S.~B.~Giddings and A.~Maharana,
  \emph{Dynamics of warped compactifications and the shape of the warped
  landscape},
  Phys.\ Rev.\  D {\bf 73} (2006) 126003
  [hep-th/0507158].

\bibitem{Baumann:2006th}
  D.~Baumann, A.~Dymarsky, I.~R.~Klebanov, J.~Maldacena, L.~McAllister and A.~Murugan,
  \emph{On D3-brane potentials in compactifications with fluxes and wrapped
  D-branes},
  JHEP {\bf 0611} (2006) 031
  [hep-th/0607050].

\bibitem{Ellis:1986pv}
  J.~R.~Ellis, C.~Gomez, D.~V.~Nanopoulos and M.~Quiros,
  \emph{World sheet instanton effects on no scale structure},
  Phys.\ Lett.\  B {\bf 173}, 59 (1986)
  [Erratum-ibid.\  {\bf 174B}, 465 (1986)].

\bibitem{Dine:1986zy}
  M.~Dine, N.~Seiberg, X.~G.~Wen and E.~Witten,
  \emph{Nonperturbative Effects on the String World Sheet},
  Nucl.\ Phys.\  B {\bf 278}, 769 (1986).

\bibitem{Dine:1987bq}
  M.~Dine, N.~Seiberg, X.~G.~Wen and E.~Witten,
  \emph{Nonperturbative Effects on the String World Sheet. 2},
  Nucl.\ Phys.\  B {\bf 289}, 319 (1987).

\bibitem{Distler:1986wm}
  J.~Distler,
  \emph{Resurrecting (2,0) Compactifications},
  Phys.\ Lett.\  B {\bf 188}, 431 (1987).

\bibitem{Gray:2007mg}
  J.~Gray, A.~Lukas and B.~Ovrut,
  \emph{Perturbative anti-brane potentials in heterotic M-theory},
  hep-th/0701025.

\bibitem{Correia:2006vf}
  F.~Paccetti Correia, M.~G.~Schmidt and Z.~Tavartkiladze,
  \emph{Radion stabilization in 5D SUGRA},
  Nucl.\ Phys.\  B {\bf 763} (2007) 247
  [hep-th/0608058].

\end{thebibliography}
\end{document}